\documentclass[superscriptaddress,twocolumn,showpacs,pra]{revtex4-1}
\usepackage[latin9]{inputenc}
\usepackage{amsmath}
\usepackage{amssymb}
\usepackage{graphicx}

\makeatletter
 
 \@ifundefined{textcolor}{}
 {%
   \definecolor{BLACK}{gray}{0}
   \definecolor{WHITE}{gray}{1}
   \definecolor{RED}{rgb}{1,0,0}
   \definecolor{GREEN}{rgb}{0,1,0}
   \definecolor{BLUE}{rgb}{0,0,1}
   \definecolor{CYAN}{cmyk}{1,0,0,0}
   \definecolor{MAGENTA}{cmyk}{0,1,0,0}
   \definecolor{YELLOW}{cmyk}{0,0,1,0}
 }

\usepackage{dcolumn}
\usepackage{bm}\usepackage{epstopdf}

\makeatother

\begin{document}

\title{Thermalization in closed quantum systems: Semiclassical approach}

\author{J. G. Cosme}

\affiliation{Institute for Advanced Study and Centre for Theoretical Chemistry
and Physics, Massey University, Auckland, New Zealand}

\author{O. Fialko}

\affiliation{Institute of Natural and Mathematical Sciences and Centre for Theoretical
Chemistry and Physics, Massey University, Auckland, New Zealand}

\date{\today}
\begin{abstract}
Thermalization in closed quantum systems can be understood either
by means of the eigenstate thermalization hypothesis or the concept
of canonical typicality. Both concepts are based on quantum mechanical
formalism such as spectral properties of the eigenstates or entanglement
between subsystems respectively. Here we study instead the onset of
thermalization of Bose particles in a two-band double well potential
using the truncated Wigner approximation. This allows us to use the
familiar classical formalism to understand quantum thermalization
in this system. In particular, we demonstrate that sampling of an
initial quantum state mimics a statistical mechanical ensemble, while
subsequent chaotic classical evolution turns the initial quantum state
into the thermal state. 
\end{abstract}
\maketitle

\section{Introduction }

Thermalization is regarded as one of the most fundamental facts in
physics. Its foundations and basic postulates are still the subject
of debates. Recent advances in addressing this issue have brought
us two quantum concepts that shed light on the onset of thermal
equilibrium inside closed quantum systems. One of them is the so-called
eigenstate thermalization hypothesis (ETH, \cite{Deutsch1991,Srednicki94}).
ETH conjectures that under certain initial conditions the expectation
values of observables of a quantum system behave as if they were thermal
during later times of its evolution. In another approach, so-called
canonical typicality (CT, \cite{Tasaki98}), the process of thermalization
is explained via entanglement between subsystems \cite{Popescu06}.
However, the quantum ideas can be hard to imagine. Semiclassical ideas,
on the other hand, may provide intuitive physical insights into quantum
mechanics \cite{Heller93}. Here we study the onset of thermalization
in a quantum system using the semiclassical truncated Wigner approximation.
This allows us to use the familiar classical formalism to explain
how quantum fluctuations in an initial state turn into thermal fluctuations
at later times during chaotic classical evolution of Wigner trajectories. 

We introduce notations to be used throughout the rest of the paper
by reviewing briefly recent progress on quantum thermalization. ETH
was tested against another hypothesis numerically in \cite{Rigol08}
(see also \cite{Rigol12}). To understand ETH, consider the initial
state $|\phi_{0}\rangle=\sum_{k}\alpha_{k}|k\rangle$, where $|k\rangle$
are the eigenstates of a Hamiltonian $\hat{H}$ with eigenvalues $E_{k}$.
The eigenstates are thermal, which is reflected in the spectral properties
as follows. The state evolves as $|\phi(t)\rangle=e^{-i\hat{H}t/\hbar}|\phi_{0}\rangle$.
As a prerequisite for thermalization, the expectation value of an
observable $\langle\hat{\mathcal{O}}\rangle=\sum_{k,l}\alpha_{l}^{\ast}\alpha_{k}e^{i(E_{l}-E_{k})t/\hbar}\mathcal{O}_{lk}$
with $\mathcal{O}_{lk}=\langle l|\hat{\mathcal{O}}|k\rangle$, after
sufficient time, must relax to the long-time average $\overline{\langle\hat{\mathcal{O}}\rangle}=\sum_{k}|\alpha_{k}|^{2}\mathcal{O}_{kk}$.
ETH then states that this occurs if $\mathcal{O}_{kk}$ is a smooth
function of $E_{k}$, while the off-diagonal elements ${\cal O}_{kl}$
are negligible. Moreover, the energy $E_{0}=\sum_{k}|\alpha_{k}|^{2}E_{k}$
must have small uncertainty $\Delta E_{0}=\sum_{k}|\alpha_{k}|^{2}(E_{k}-E_{0})^{2}$
\cite{Srednicki94,Rigol08} to ensure that the relaxed state is thermal
and depends only on the energy. 

CT, on the other hand, replaces the need for any ensemble averaging,
the main postulate of statistical mechanics considered to be artificial.
Here, thermalization is reached on the level of a subsystem. The reduced
density matrix of the subsystem is canonical for the majority of the possible
pure states of the entire many-body system under global constraint
such as energy. This was established under great generality by invoking
Levy's lemma \cite{Popescu06}. Then, the properties of the subsystem
can be calculated using the reduced density matrix constructed from
the pure density matrix of the entire system $\hat{\rho}_{p}=|\phi(t)\rangle\langle\phi(t)|$
giving the same prediction as if the entire system was in the microcanonical
state $\hat{\rho}_{m}=\mathcal{N}_{E_{0},\delta}^{-1}\sum_{k}|k\rangle\langle k|$.
The sum is over $\mathcal{N}_{E_{0},\delta}$ eigenstates $|k\rangle$,
with energies lying within some window $[E_{0}-\delta,E_{0}+\delta]$
such that $\delta\ll E_{0}$.

In classical physics, thermalization is explained by means of classical
chaos. Being chaotic, the system wanders all over the constant energy
surface in the phase space, becoming ergodic. Averaged properties of
the system over a long time can then be estimated by averaging over
the accessible phase space \cite{Reif85}. To relate that to thermalization
in closed quantum systems, it was originally argued that ETH can be
explained if the quantum system is chaotic in the classical limit
\cite{Srednicki94}. However, many quantum systems do not have classical
counterparts; nevertheless, they thermalize \cite{Rigol08,Denisov,Steinigeweg14}.
Thermalization in such systems is believed to be related to the onset
of chaotic eigenstates in the system \cite{Flambaum97 ,Santos12}.
 To demonstrate this. we write the density matrix corresponding to
$|\phi(t)\rangle$ as

\begin{equation}
\hat{\rho}_{p}=\sum_{k}|\alpha_{k}|^{2}|k\rangle\langle k|+\sum_{k\ne l}\alpha_{k}\alpha_{l}^{\ast}e^{-i(E_{k}-E_{l})t/\hbar}|k\rangle\langle l|.\label{eq:CT}
\end{equation}
Here the coefficients $\alpha_{k}$ are assumed to be random and independent,
implying that the second term quickly averages to zero at long times.
In the remaining first term, the factor $|\alpha_{k}|^{2}$ is assumed
to be a smooth function with narrow width $\Delta E_{0}\ll E_{0}$,
yielding $\hat{\rho}_{p}\approx\hat{\rho}_{m}$.

\section{The model}

We study thermalization in a quantum system which is amenable to the
semiclassical analysis. Bosons are trapped in a double-well potential
shown in Fig. \ref{fig:Model}. They occupy four energy levels (we
will refer to them also as modes) and are described by the following
two-band Bose-Hubbard Hamiltonian \cite{Fialko12}: 
\begin{figure}[!b]
 \includegraphics[width=0.8\columnwidth]{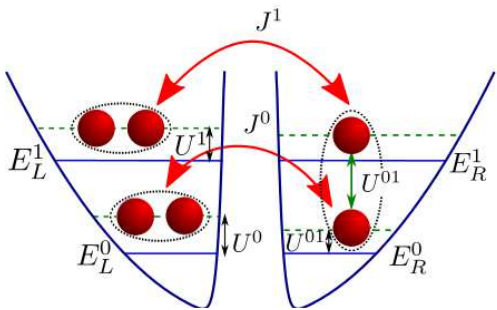} \protect\caption{Schematic of the double-well potential with two energy bands. The
diagram shows the tunneling of particles and how the energy levels
change due to the interactions between them. The interband coupling
$U^{01}$ makes the system complex enough to show quantum thermalization
of the particles. \label{fig:Model}}
\end{figure}
 
\begin{align}
\hat{H} & =-\sum_{r\neq r',l}J^{l}\hat{b}_{r}^{l\dagger}\hat{b}_{r'}^{l}+\sum_{r,l}U^{l}\hat{n}_{r}^{l}(\hat{n}_{r}^{l}-1)+\sum_{r,l}E_{r}^{l}\hat{n}_{r}^{l}\nonumber \\
 & +U^{01}\sum_{r,l\neq l'}(2\hat{n}_{r}^{l}\hat{n}_{r}^{l'}+\hat{b}_{r}^{l\dagger}\hat{b}_{r}^{l\dagger}\hat{b}_{r}^{l'}\hat{b}_{r}^{l'}),\label{eq:Hamilt}
\end{align}
where $\hat{b}_{r}^{l\dagger}$ and $\hat{b}_{r}^{l}$ are the bosonic
creation and annihilation operators respectively of an atom in well
$r$ and energy level $l$. The parameters in the Hamiltonian can
be easily evaluated for a specific double-well potential \cite{Fialko12}.
The ground and first excited-state energies are $E_{r}^{l}=\int dx\phi_{r}^{l*}(x)\hat{H}_{sp}\phi_{r}^{l}(x)$.
The tunneling term between wells is $J^{l}=-\int dx\phi_{L}^{l*}(x)\hat{H}_{sp}\phi_{R}^{l}(x)$.
The interaction term between atoms in the same well and on the same
energy level is $U^{l}=g\int dx|\phi_{r}^{l}|^{4}$ and on different
energy levels is $U^{01}=g\int dx|\phi_{r}^{0}(x)|^{2}|\phi_{r}^{1}(x)|^{2}$.
This term contributes to atoms changing energy levels. We consider
a harmonic potential with oscillator frequency $\omega_{0}$, which
is split by a focused laser beam located at the center of the trap
and described by a Gaussian potential $V_{0}\mathrm{exp}(-x^{2}/2\sigma^{2})$.
The barrier height is chosen to be $V_{0}=5\hbar\omega_{0}$ with
width $\sigma=0.1l_{ho}$, where the harmonic oscillator length is
$l_{ho}=\sqrt{\hbar/m\omega_{0}}$. The localized functions $\phi_{r}^{l}$
are obtained by numerically solving the eigenstates of the single particle Hamiltonian $\hat{H}_{sp}$.
The coupling $g$ can be varied by Feshbach resonance in an experiment.
Parameters in units of the harmonic confinement $\hbar\omega_{0}$
are $J^{0}=0.26$, $J^{1}=0.34$, $E_{r}^{0}=1.25$, $E_{r}^{1}=3.17$,
$U^{0}=4/N$, $U^{1}=3U^{0}/4$, and $U^{01}=U^{0}/2$. For $N=40$ particles,
the system is complex enough to show thermalization and can be studied
by exact diagonalization. A small number of modes and large number of
particles allow us to study the system also in the semiclassical limit.

\section{Truncated Wigner approximation}

For large $N$, we can use the semiclassical truncated Wigner approximation
(TWA). The leading corrections due to finiteness of $N$ is of the order
$1/N^{2}$ \cite{Polkovnikov10}. For $N=40$ we can safely ignore
such terms and the results can be compared to exact diagonalization
(see below). The use of TWA is justified since the initial state sampling
naturally mimics the ensemble averaging in statistical mechanics.
As a matter of fact, it was conjectured that it is probable that each
Wigner trajectory approximately corresponds to a single realization
of experiments \cite{Polkovnikov10}. Within TWA the operators are
treated as complex numbers, ${b}_{r}^{l}$ and ${b}_{r}^{l\ast}$,
satisfying the following set of nonlinear equations: 
\begin{equation}
i\hbar\frac{\partial{b}_{r}^{l}}{\partial t}=\frac{\partial H_{W}}{\partial{b}_{r}^{l*}},\label{semi}
\end{equation}
where $H_{W}=\langle\beta|\hat{H}|\beta\rangle$ is the Weyl-ordered
Hamiltonian operator. It is calculated by replacing $\hat{b}_{r}^{l}\to\biggl({b}_{r}^{l}+\frac{1}{2}\frac{\partial}{\partial{b}_{r}^{l*}}\biggr)$
and $\hat{b}_{r}^{l\dagger}\to\biggl({b}_{r}^{l*}-\frac{1}{2}\frac{\partial}{\partial{b}_{r}^{l}}\biggr)$
in Eq. (\ref{eq:Hamilt}) \cite{Polkovnikov10}. We solve Eq. (\ref{semi})
by sampling appropriately an initial quantum state. If the initial
state is a Fock state, $b_{r}^{l}$ are sampled for large $N$ with
fixed amplitude $\sqrt{n_{r}^{l}}$ and uniformly random phase $\theta$,
i.e., $b_{r}^{l}=\sqrt{n_{r}^{l}}\exp(i\theta)$ \cite{Olsen09}. We
also study dynamics from the coherent state sampled as $b_{r}^{l}=\sqrt{n_{r}^{l}}+1/2(\eta_{1}+i\eta_{2})$,
where $\eta_{j}$ are real normal Gaussians. These have the correlations
$\overline{\eta_{j}}=0$ and $\overline{\eta_{j}\eta_{k}}=\delta_{jk}$,
where the overline denotes an average over many samples \cite{Olsen09}.
The occupation numbers can be calculated by averaging $|b_{r}^{l}(t)|^{2}$
over many realizations from the initial sampling. To ensure convergence
the number of trajectories and number of particles are taken to be
large, $N=10^{4}$. 

The dynamics of ultracold atoms in a single-band double well potential
has been studied extensively \cite{Double-well}. This system can
be mapped onto the classical pendulum with small Josephson oscillations
and the self-trapping regime of atoms being identified with small-amplitude
oscillations and full rotation of the pendulum around its pivot point,
respectively. Similarly, a two-band double-well potential filled with
cold atoms can be described as two nontrivially coupled nonrigid pendulums
in the semiclassical limit \cite{pendula}. As a result, there are
regimes, where the system is chaotic \cite{CMbook}. With our choice
of parameters, the Wigner trajectories are indeed chaotic as shown
below. 

Thermal equilibration through chaotic semiclassical dynamics has been
studied before, e.g., in the Dicke model \cite{dicke12} and in trapped
atomic gases with spin-orbit coupling \cite{Larson13}. In this paper
we extend that by showing in detail how an initial quantum state turns
into the thermal state through chaotic evolution. The proper sampling
of an initial quantum state mimics a statistical mechanical ensemble,
while ergodicity of Wigner trajectories drives the initial pure distribution
to the thermal one. 
\begin{figure}[!t]
\includegraphics[width=0.9\columnwidth]{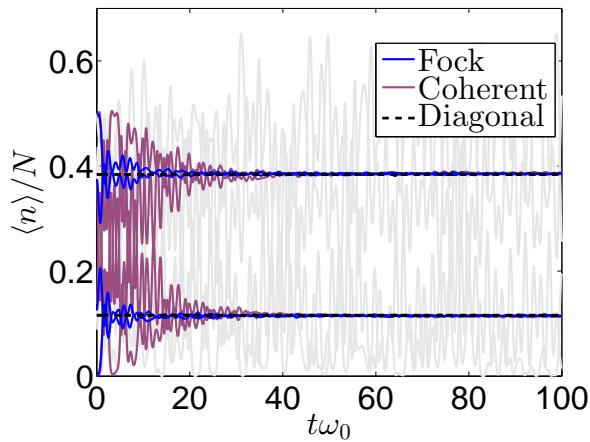} \protect\caption{TWA dynamics of the occupation numbers in each mode from Fock and
coherent initial states with the same energy $E_{0}/N\approx2.25\hbar\omega_{0}$
and narrow energy variances $\Delta E_{0}/E_{0}=0.09$ and $0.02$
respectively. They equilibrate at values which are in excellent
agreement with the quantum diagonal ensembles prediction for $N=40$.
A single TWA trajectory is shown in semitransparent gray; it exhibits
chaotic behavior. Only averaging over many such trajectories leads
to the correct relaxed values of the occupation numbers.}

\label{twadiagmic-1} 
\end{figure}

\section{Exact diagonalization }

The semiclassical analysis will be compared with the predictions of
the full quantum dynamics for consistency. We use the Fock basis $|n\rangle=|n_{L}^{0}\rangle\otimes|n_{R}^{0}\rangle\otimes|n_{L}^{1}\rangle\otimes|n_{R}^{1}\rangle$,
where $|n_{r}^{l}\rangle$ is a number state on level $l$ and site
$r$. The eigenstates are $|k\rangle=\sum_{n}C_{n}^{k}|n\rangle$,
where $C_{n}^{k}$ are extracted from exact diagonalization of the
Hamiltonian (\ref{eq:Hamilt}) for $N=40$ particles. Assume that
the initial state $|\phi_{0}\rangle=|n_{0}\rangle$ is a Fock state
with fixed energy $E_{0}$, implying that the coefficients $\alpha_{k}\equiv C_{n_{0}}^{k\ast}$
are obtained from the exact diagonalization (ED). The details of the
initial state are irrelevant for subsequent evolution if its energy
variance is small, as we discussed above. The system for any such initial
state with the same energy $E_{0}$ relaxes to the diagonal ensemble
$\hat{\rho}_{m}=\sum_{k}|C_{n_{0}}^{k}|^{2}|k\rangle\langle k|$. 

Another commonly used state albeit quite different from the Fock state,
is a coherent state, which is a superposition of all possible Fock
states, $|\beta\rangle=e^{-|b|^{2}/2}\sum_{n=0}^{\infty}(b^{n}/\sqrt{n!})|n\rangle$.
Inserting the resolution of identity expressed via the Fock basis, we
get $\langle\beta|\hat{H}|\beta\rangle=\sum_{n,m}\langle\beta|n\rangle\langle m|\beta\rangle E_{nm}$,
where $E_{nm}=\langle n|\hat{H}|m\rangle$. We split it into the sums
over diagonal and off-diagonal terms, $\langle\beta|\hat{H}|\beta\rangle=\sum_{n}e^{-|b|^{2}}\frac{|b|^{2n}}{n!}E_{n}+\sum_{n\neq m}e^{-|b|^{2}}\frac{b^{*n}b^{m}}{\sqrt{n!m!}}E_{nm}.$
Since we require the energy variance in the Fock basis to be small,
which translates to $\sqrt{\sum_{m\neq n}E_{nm}^{2}}\ll E{}_{n}$,
the off-diagonal elements in the Hamiltonian matrix $E_{nm}$ are
negligibly small as compared to the diagonal elements $E_{n}$. Therefore
we neglect the terms with $n\neq m$. For a quantum system with large
number of particles in each mode, $|b|^{2}\gg1$, the Poissonian factor
$e^{-|b|^{2}}\frac{|b|^{2n}}{n!}$ becomes sharply peaked at around $n=|b|^{2}$.
This allows us to approximate $\langle\beta|\hat{H}|\beta\rangle\approx\langle n|\hat{H}|n\rangle$
with $n=|b|^{2}$. The energy variance of the coherent state can be
shown to get progressively small when we increase the number of particles.
Therefore, it is supposed to relax to the diagonal ensemble $\hat{\rho}_{m}$
if it has the same energy $E_{0}$. We examine this within the semiclassical
approach. 
\begin{figure}[!t]
\includegraphics[width=0.5\columnwidth]{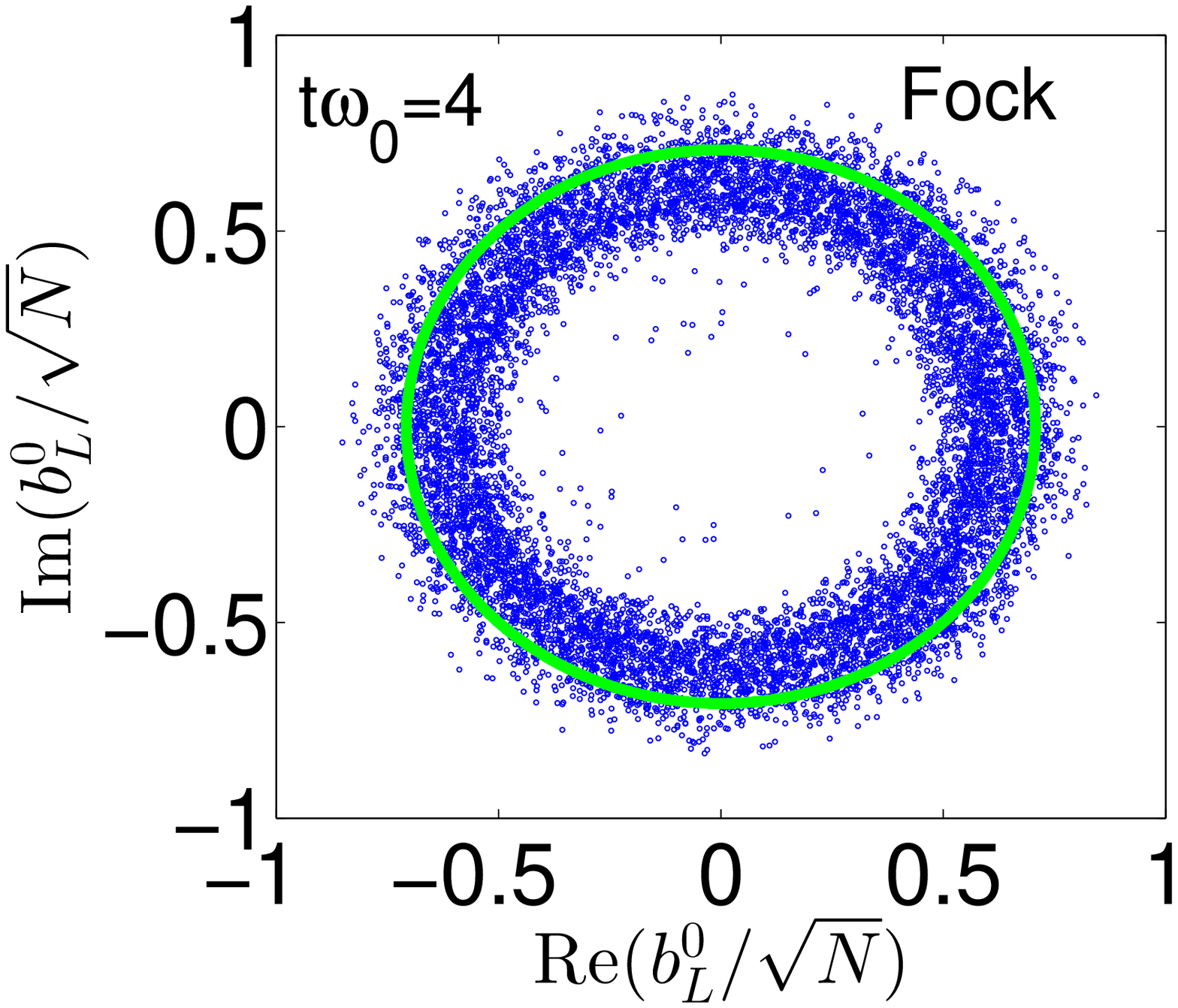}\includegraphics[width=0.5\columnwidth]{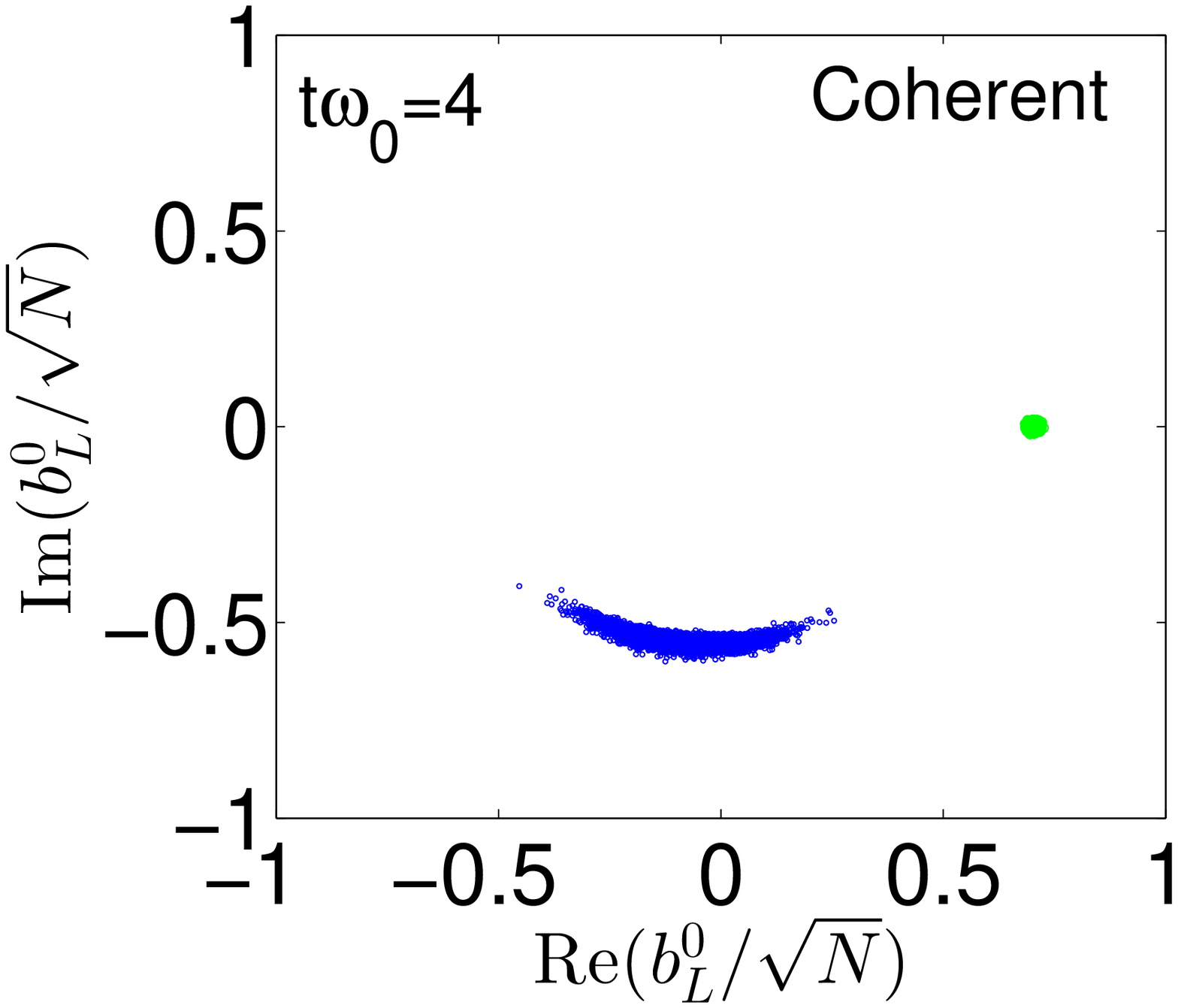}

\includegraphics[width=0.5\columnwidth]{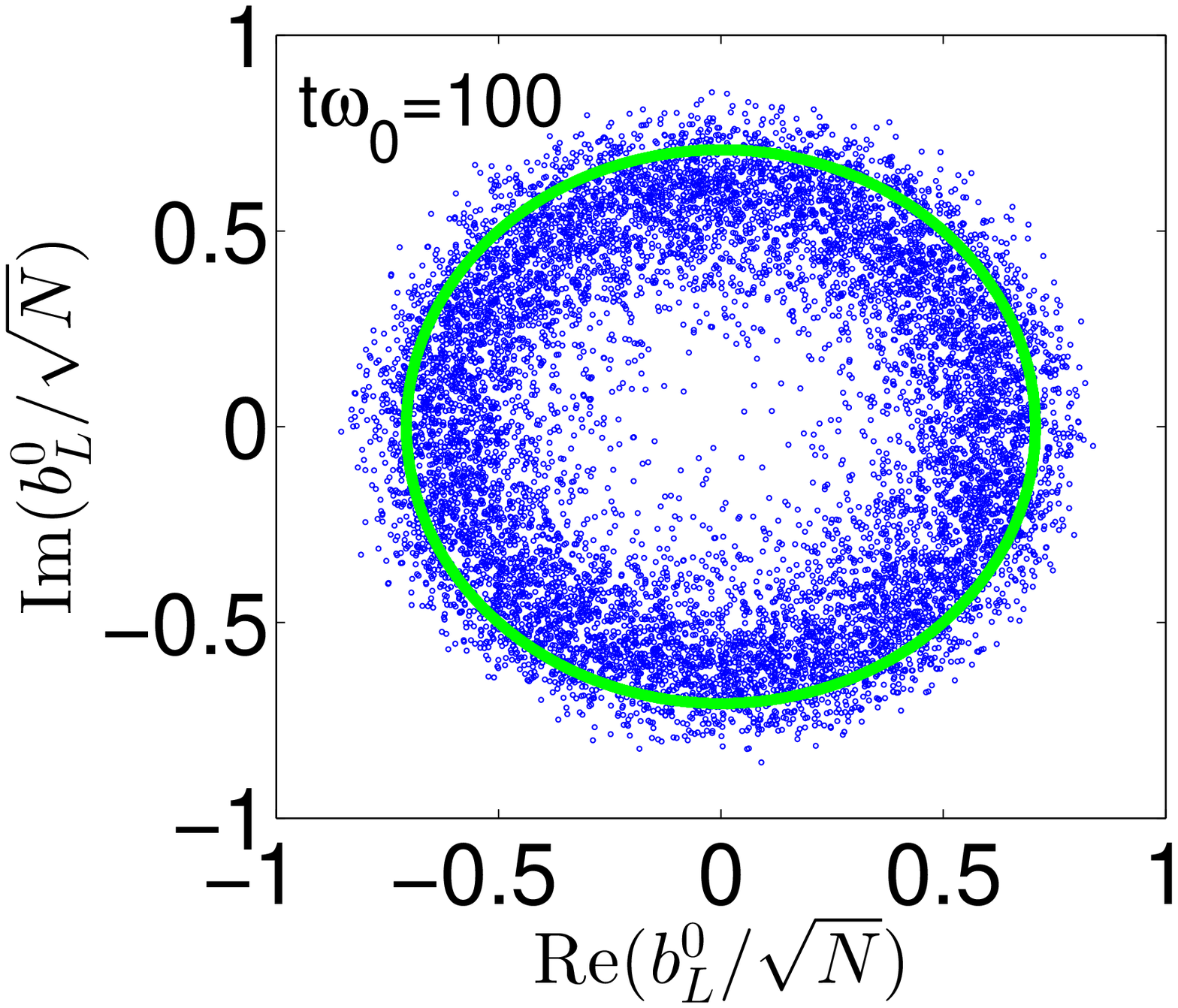}\includegraphics[width=0.5\columnwidth]{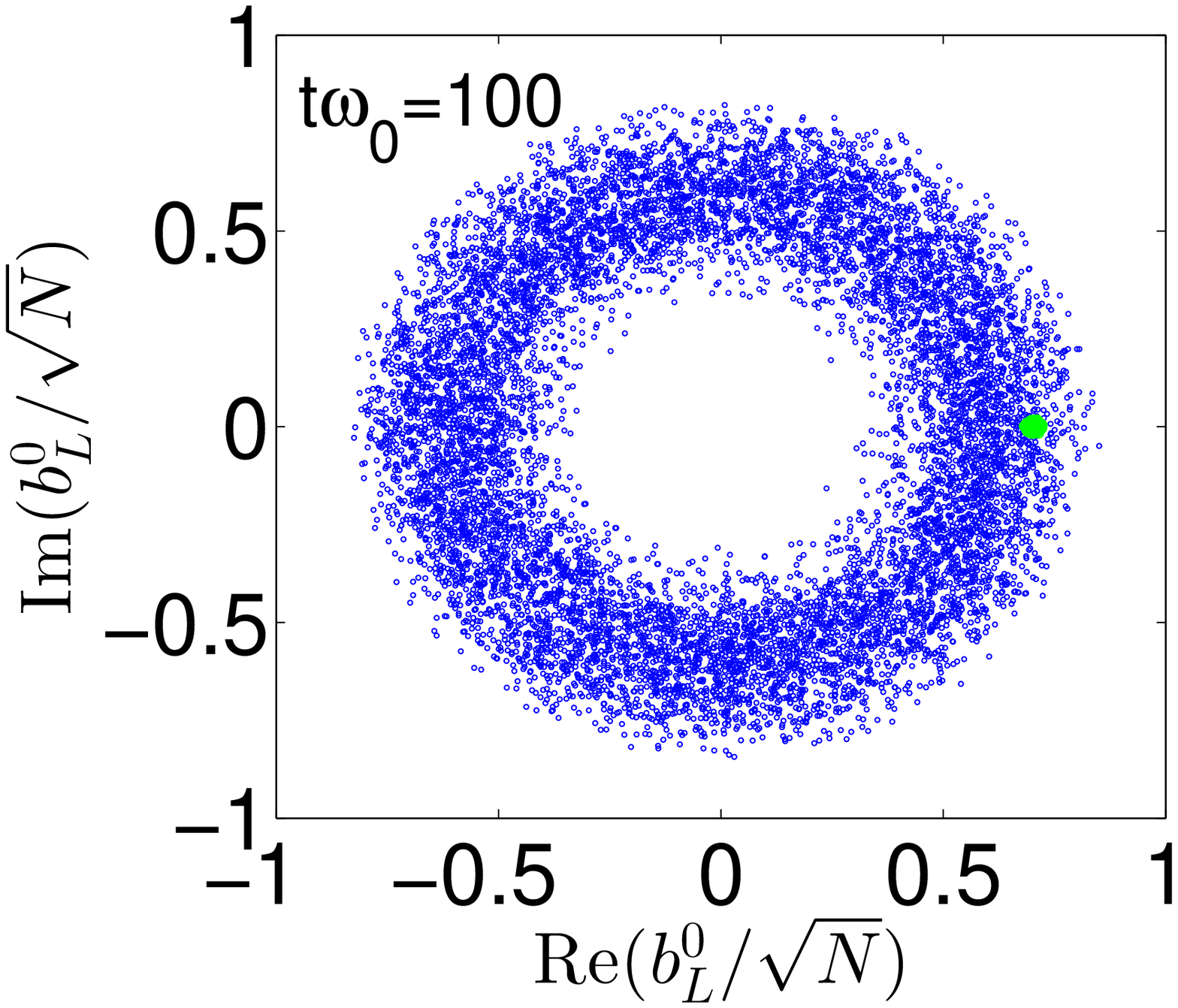}

\protect\caption{Ergodicity in TWA. As an example we show Wigner distributions of one
of the lower levels. Initially the system is prepared in the Fock state
(green cycle shown on the left) or in the coherent state (green bump
shown on the right). Trajectories sampled from the initial states
fill the available phase space as time evolves (blue dots). The distributions
become essentially the same at some time and after that they do not
change, which suggests thermalization is reached within TWA. In this
sense, quantum fluctuations of the initial state turn into thermal
fluctuations in the course of evolution. }

\label{bdist01} 
\end{figure}

\section{Results}

To compare ED and TWA, we scale all quantities with $N.$ We study
the dynamics from initial Fock and coherent states with the same energy
$E_{0}/N\approx2.25\hbar\omega_{0}$. The relaxation dynamics from
TWA calculations is shown in Fig. \ref{twadiagmic-1}. The final relaxed
values of the occupation numbers are in excellent agreement with the
quantum diagonal ensembles prediction. For the two different initial
conditions the upper and lower modes thermalize at $n_{L}^{0}\approx n_{R}^{0}\approx0.4N,n_{L}^{1}\approx n_{R}^{1}\approx0.1N$.
The independence on initial conditions is the hallmark of thermalization.
The Wigner trajectories exhibit chaotic behavior by quickly filling
the available phase space, as shown in Fig. \ref{bdist01}. This is
quite similar to classical thermalization, where such behavior of
classical trajectories is the source of ergodicity in an ensemble
of identical systems \cite{Reif85}. The quantum-mechanical fuzziness
is the key to understand ergodicity in the semiclassical language:
averaging over the initial sampling replaces  ensemble averaging. 

\begin{figure}[!t]
\includegraphics[width=0.47\columnwidth]{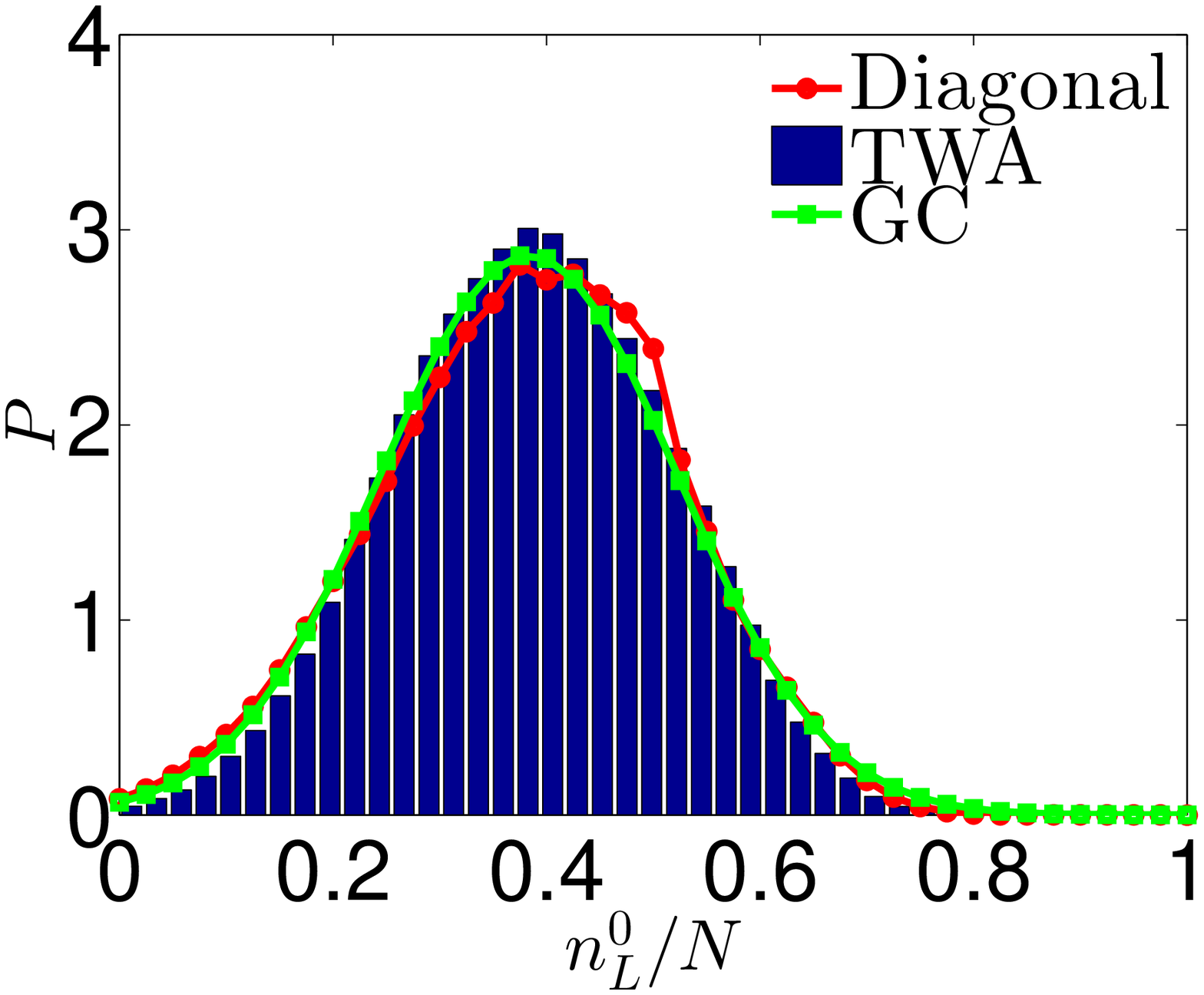} \includegraphics[width=0.47\columnwidth]{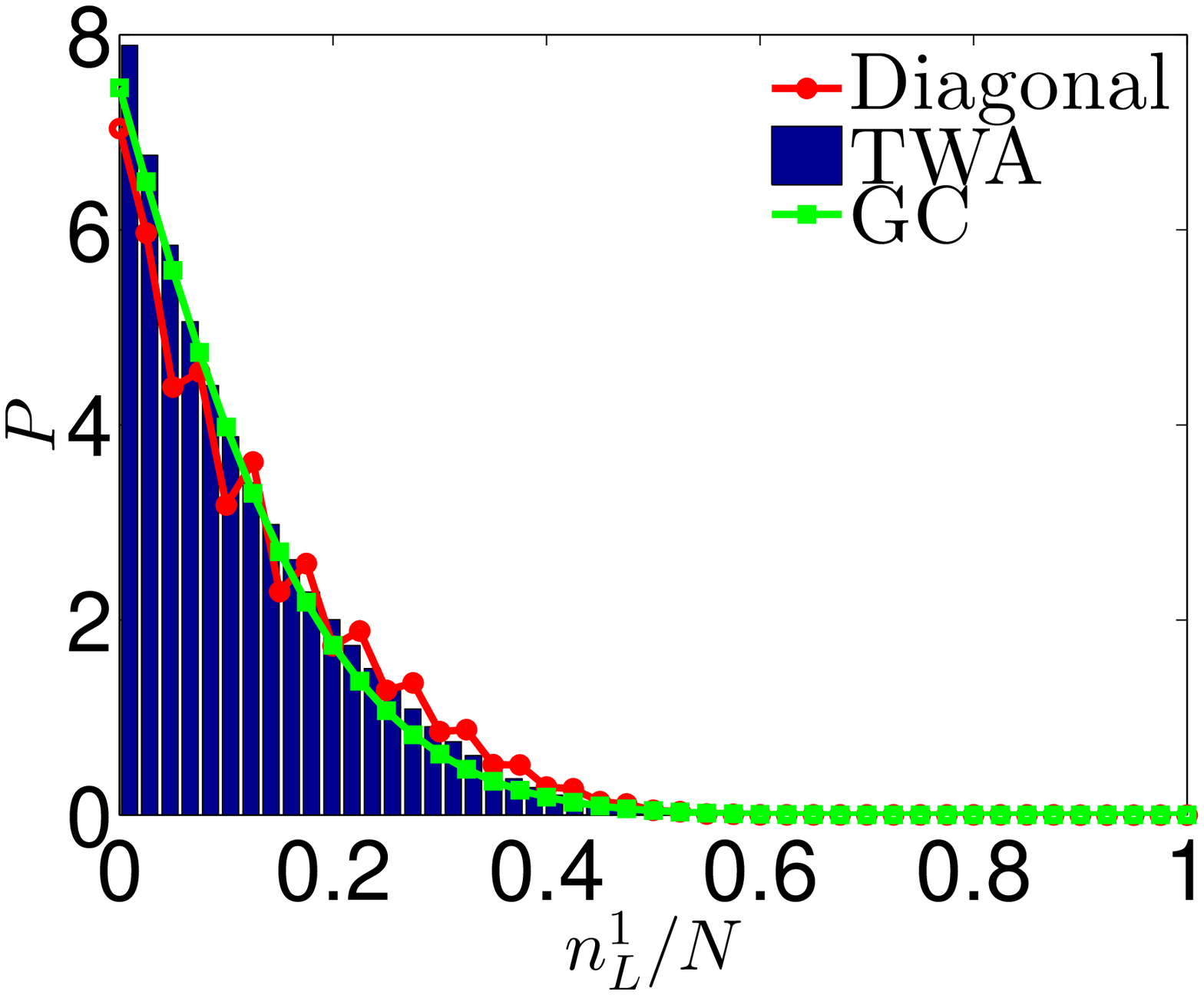}

\includegraphics[width=0.47\columnwidth]{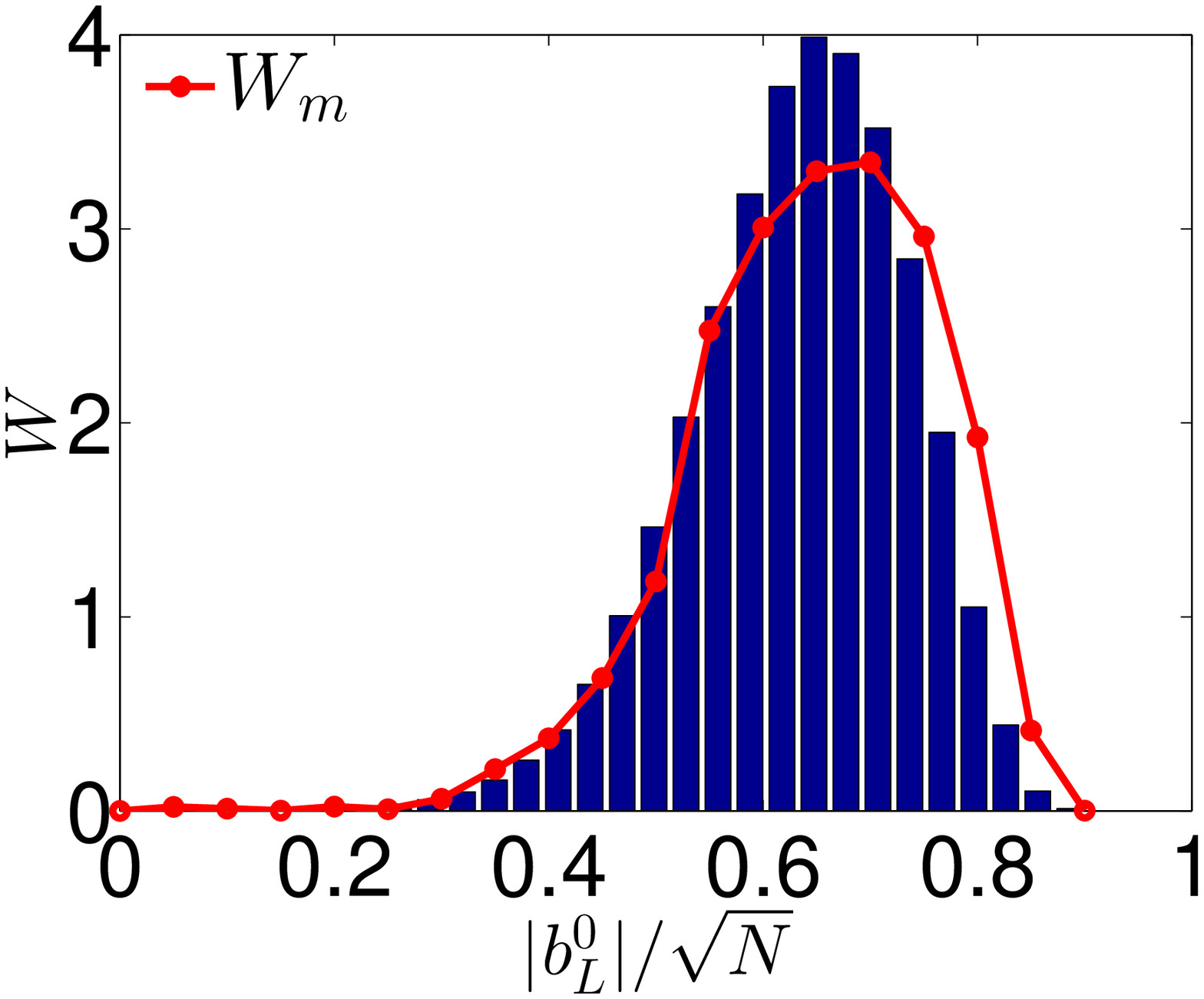}\includegraphics[width=0.47\columnwidth]{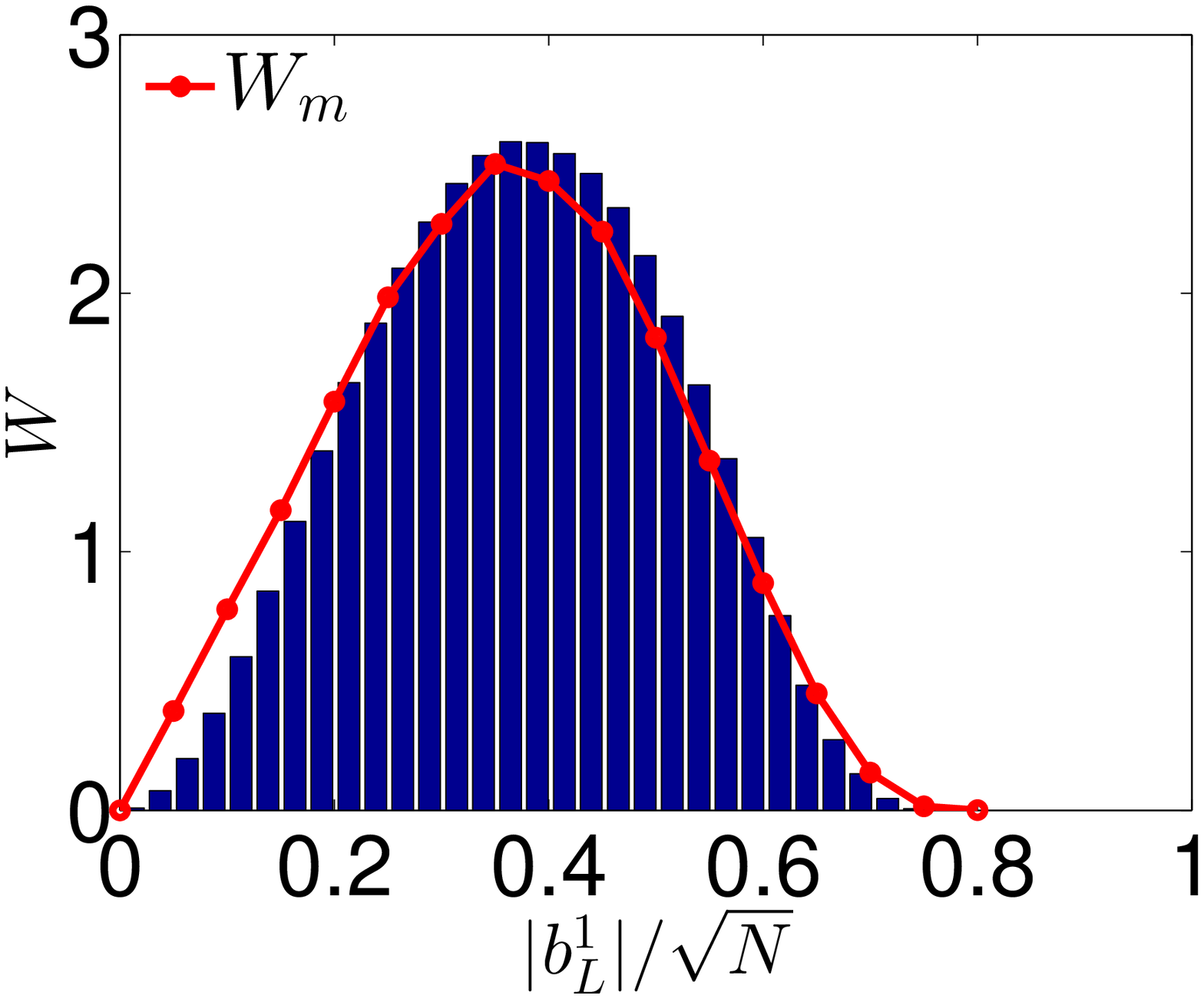}

\protect\caption{(Upper panels) Comparison of the distributions $P_{n}$ derived from
the diagonal ensemble with the corresponding distributions derived
from the grand-canonical ensembles. Both are compared with the corresponding
data extracted from TWA. (Lower panels) The good agreement between
the exact Wigner distribution and the distribution $W_{m}$.}

\label{deltvstwa} 
\end{figure}

Within the realm of ED the diagonal ensemble $\hat{\rho}_{m}$ contains
all necessary information about the relaxed state. Our aim is to calculate
the population distribution in a mode $P_{n}$ and compare it with
the analogous distribution derived from the TWA approximation. This quantity
simply gives the occupancy of that mode, $\langle\hat{n}\rangle=\sum_{n'}n'P_{n'}$,
and the reduced density matrix of that mode, $\sum_{n'}P_{n'}|n'\rangle\langle n'|$.
To calculate this density distribution we express $\hat{\rho}_{m}$
via the Fock basis $\hat{\rho}_{m}\approx\sum_{k,n}|C_{n_{0}}^{k}|^{2}|C_{n}^{k}|^{2}|n\rangle\langle n|$,
where again we rely on the chaoticity of the eigenstates, leaving only
the diagonal contribution, yielding $P_{n}=\sum_{k}|C_{n_{0}}^{k}|^{2}|C_{n}^{k}|^{2}$.
We found a very good agreement with the analogous distribution obtained
from the TWA calculations, as it is shown in Fig. \ref{deltvstwa}.
The later is obtained by noting that $n_{r}^{l}\rightarrow|b_{r}^{l}|^{2}$. 

The resulting population distributions can be inferred from the ergodicity
of the Wigner trajectories originating from the initial sampling which
is narrow in energy. The Wigner distribution function of the entire
system can thus be represented as some sharply peaked function $\bar{\delta}(H_{W}-E_{0})$,
the microcanonical analog of the quantum case. We approximate it
with the Gaussian $\bar{\delta}(x)=\frac{1}{\sigma\sqrt{2\pi}}\mathrm{exp}(-x^{2}/2\sigma^{2})$
with $\sigma\sim\Delta E_{0}$. The Wigner distribution of a mode
$b_{r}^{l}$ is then given by integrating out over all modes but one,
$W_{m}(b_{r}^{l},b_{r}^{l*})=\int\bar{\delta}(H_{W}-E_{0})\prod_{l'\neq l,~r'\neq r}{db_{r'}^{l'}}^{*}db_{r'}^{l'}$,
which is numerically evaluated using the Monte Carlo integration.
We found good agreement with the exact Wigner distributions as shown
in Fig. \ref{deltvstwa}. This is conceptually analogous to CT, although
the underlying basis of our approach is quite distinct.

Having observed equilibration in each of the four modes, we now examine
the states of each mode in more detail. Our aim is to demonstrate
that the equilibrium states are thermal. In studying thermalization
in closed quantum systems, the Hamiltonian of a system is usually
split into several parts, $\hat{H}=\hat{H}_{S}+\hat{H}_{B}+\hat{H}_{int}$,
representing the subsystem, the bath, and interactions between them
leading to their mutual thermalization. Equation (\ref{eq:Hamilt}) represents
such a situation: Each of the four modes can be regarded as a subsystem
($=\hat{H}_{S}$) coupled to the rest of the system ($=\hat{H}_{B})$,
via the tunneling and coupling terms $(=\hat{H}_{int}$). The modes
may exchange energy and particles via tunneling; therefore thermal
states of each mode are expected to be described by grand-canonical
ensembles $\hat{\rho}_{GC}\sim e^{-\beta(\hat{H}_{S}-\mu\hat{n}_{S})}$,
where for a given mode $\hat{H}_{S}=U^{l}\hat{n}_{r}^{l}(\hat{n}_{r}^{l}-1)+E_{r}^{l}\hat{n}_{r}^{l}$
and $\hat{n}_{S}=\hat{n}_{r}^{l}$. To ensure that the resulting distributions
are thermal, we fitted them with the grand-canonical distribution
$\hat{\rho}_{GC}$. We have extracted $\beta^{-1}\approx6\hbar\omega_{0}$,
$\mu\approx3\hbar\omega_{0}$ and $\beta^{-1}\approx12\hbar\omega_{0}$,
$\mu\approx0.25\hbar\omega_{0}$ for the lower and upper modes, respectively.
In passing, we note that if we chose, e.g., the inter-band coupling $U^{01}$
to be much smaller than the current one, the situation is different:
the equilibration is lost along with the chaotic behavior of the Wigner
trajectories.

\section{Discussions and Conclusions }

We have demonstrated that each of the four modes thermalizes with
the rest of the system. We did this as follows. On the one hand, each
mode is weakly coupled to each other via tunneling and interaction
terms in the Hamiltonian so that they can exchange particles and energy.
As a result, the corresponding distributions are expected (and were
shown) to be well described by the grand-canonical ensembles with
appropriate temperatures and chemical potentials. On the other hand,
the initial state independence of the final states provides further
evidence of thermalization of each mode. Moreover, the final states
of each mode can be inferred from the microcanonical ensemble of
the whole system. 

We have also shown that the semiclassical truncated Wigner approach
and full quantum description agree in reproducing the states of each
mode after they have been thermalized. While the quantum description
of thermalization has been elucidated extensively in the literature
\cite{Deutsch1991,Srednicki94,Tasaki98,Popescu06,Rigol12}, in this
work we analyzed the semiclassical approach in order to seek further
insights into the physics of quantum thermalization. The truncated
Wigner approach has revealed deep connection between thermalization
in closed quantum systems and classical thermalization. Although the
main postulate of classical statistical mechanics is considered to
be artificial and was replaced in quantum formalism \cite{Popescu06},
we have shown that quantum-mechanical fuzziness of initial states
within the semiclassical formalism naturally supports the concept
of statistical mechanical ensembles. Moreover, the ergodicity of Wigner
trajectories leads to thermal relaxation. Being conceptually different,
our study adds to the understanding of quantum thermalization and
to the recent advances in pushing the limits of quantum thermalization
and its understanding in the macroscopic limit \cite{Steinigeweg14,Rigol14,Larson13,dicke12}. 
\begin{acknowledgments}
We thank Anatoli Polkovnikov and Joachim Brand for useful and insightful
comments. O.F. was supported by the Marsden Fund (Project No. MAU1205), administrated
by the Royal Society of New Zealand.\end{acknowledgments}


\begin{thebibliography}{10}
\bibitem{Deutsch1991} J. M. Deutsch, Phys. Rev. A \textbf{43}, 2046
(1991).

\bibitem{Srednicki94}M. Srednicki, Phys. Rev. E \textbf{50}, 888
(1994).

\bibitem{Tasaki98}H. Tasaki, Phys. Rev. Lett. \textbf{80}, 1373 (1998)
; S. Goldstein, J.L. Lebowitz, R. Tumulka and N. Zanghi, ibid. \textbf{96}, 050403 (2006).

\bibitem{Popescu06}S. Popescu, A.J. Short and A. Winter, Nature Phys.
\textbf{2}, 754 (2006). 

\bibitem{Heller93}E. J. Heller and S. Tomsovic, Phys. Today \textbf{46},
38 (1993).

\bibitem{Rigol08}M. Rigol, V. Dunjko and M. Olshanii, Nature (London) \textbf{452},
854 (2008); A. C. Cassidy, C. W. Clark and M. Rigol, Phys. Rev. Lett.
\textbf{106}, 140405 (2011). 

\bibitem{Rigol12}M. Rigol and M. Srednicki, Phys. Rev. Lett. \textbf{108},
110601 (2012).

\bibitem{Reif85}F. Reif, \textit{Fundamentals of Statistical and
Thermal Physics} (McGraw-Hill Book Co., New York, 1985).

\bibitem{Denisov}A. V. Ponomarev, S. Denisov and P. Hanggi, Phys.
Rev. Lett. \textbf{106}, 010405 (2011); A. V. Ponomarev, S. Denisov,
P, Hanggi and J. Gemmer, EPL \textbf{98}, 40011 (2012). 

\bibitem{Steinigeweg14}C. Ates, J.P. Garrahan and I. Lesanovsky,
Phys. Rev. Lett. \textbf{108}, 110603 (2012); R. Steinigeweg, A. Khodja,
H. Niemeyer, C. Gogolin and J. Gemmer, ibid. \textbf{112},
130403 (2014).

\bibitem{Flambaum97 }V. V. Flambaum and F. M. Izrailev, Phys. Rev.
E \textbf{56}, 5144 (1997).

\bibitem{Santos12}L. F. Santos, F. Borgonovi and F. M. Izrailev,
Phys. Rev. Lett. \textbf{108}, 094102 (2012).

\bibitem{Fialko12}M.A. Garcia-March, D.R. Dounas-Frazer and L. D.
Carr, Front. Phys. \textbf{7}, 131 (2012); O. Fialko and D.W. Hallwood,
Phys. Rev. Lett. \textbf{108}, 085303 (2012); O. Fialko, J. Phys.
B \textbf{47}, 045302 (2014).

\bibitem{Polkovnikov10}P.B. Blakie, A.S. Bradley, M.J. Davis, R.J.
Ballagh and C.W. Gardiner, Adv. Phys. \textbf{57}, 363 (2008);
A. Polkovnikov, Ann. Phys. \textbf{325}, 1790 (2010).

\bibitem{Olsen09}M.K. Olsen and A.S. Bradley, Opt. Commun. \textbf{282},
3924 (2009). 

\bibitem{Double-well}A. Smerzi, S. Fantoni, S. Giovanazzi and S.
R. Shenoy, Phys. Rev. Lett. \textbf{79}, 4950 (1997); G. J. Milburn,
J. Corney, E. M. Wright and D. F. Walls, Phys. Rev. A \textbf{55},
4318 (1997); M. Albiez, R. Gati, J. Folling, S. Hunsmann, M. Cristiani
and M. K. Oberthaler, Phys. Rev. Lett. \textbf{95}, 010402 (2005);
S. Levy , E. Lahoud, I. Shomroni and J. Steinhauer, Nature (London) \textbf{449},
579 (2007).

\bibitem{pendula} J. Gillet, M. A. Garcia-March, Th. Busch and F.
Sols, Phys. Rev. A \textbf{89}, 023614 (2014).

\bibitem{CMbook}T. W.B. Kibble and F. H. Berkshire, \textit{Classical
Mechanics} (Imperial College Press, London, 2012).

\bibitem{dicke12}A. Altland and F. Haake, New J. Phys. \textbf{14},
073011 (2012).

\bibitem{Larson13}J. Larson, B. M. Anderson and A. Altland, Phys.
Rev. A 87, 013624 (2013).

\bibitem{Rigol14}M. Rigol, Phys. Rev. Lett. \textbf{112}, 170601
(2014).\end{thebibliography}
\end{document}